# Scheduling on Grid with communication Delay

Difrawi Samouriq, gscholar4@gmail.com

International Institute of Reserach

# Abstract

Parallel processing, the core of High Performance Computing (HPC), was and still the most effective way in improving the speed of computer systems. For the past few years, the substantial developments in the computing power of processors and the network speed have strikingly changed the landscape of HPC. Geography distributed heterogeneous systems can now cooperate and share resources to execute one application. This computing infrastructure is known as computational Grid or Grid Computing. Grid can be viewed as a distributed large-scale cluster computing. From other perspective, it constitutes the major part of Cloud Computing Systems in addition to thin clients and utility computing [1,2, 3]. Hence, Grid computing has attracted many researchers [4]. The interest in Grid computing has gone beyond the paradigm of traditional Grid computing to a Wireless Grid computing [5,6].

# Introduction

Parallel processing, the core of High Performance Computing (HPC), was and still the most effective way in improving the speed of computer systems. For the past few years, the substantial developments in the computing power of processors and the network speed have strikingly changed the landscape of HPC. Geography distributed heterogeneous systems can now cooperate and share resources to execute one application. This computing infrastructure is known as computational Grid or Grid Computing. Grid can be viewed as a distributed large-scale cluster computing. From other perspective, it constitutes the major part of Cloud Computing Systems in addition to thin clients and utility computing [1,2, 3]. Hence, Grid computing has attracted many researchers [4]. The interest in Grid computing has gone beyond the paradigm of traditional Grid computing to a Wireless Grid computing [5,6].

One challenging problem in Grids Computing is finding an analytical performance Scheduling model that considers the communication speed and the processing speed at the same time. A well-known tool for this purpose is the stochastic queuing theory [7, 8, 9, 10]. Another important tool is using Divisible Load Theory (DLT)[31]. In principle divisible load theory is a deterministic theory; however, it has been shown that for great instance it is equivalent to Markov Chain Modeling [11].

DLT has demonstrated its effectiveness in scheduling and allocating very large independent tasks on Grid originated from multiple resources [12-14]. This work does not consider the communication time or consider it negligible, and so it was not included in the analytical model developed. In [36] Communication time is considered; however, the finish time did not have a closed form solution. In [8] communication time is well thought-out but not in dividing the load so the transfer input time of the load was not part of the model.

In this work, we propose an analytical scheduling model for the Grid based on DLT, which takes into considerations the communication time as well as the computation time.

## Literature Review

The most important and inherent problem in traditional distributed systems such as loosely coupled parallel systems, massively parallel processors computers(MPP), cluster of workstations(COW) and, of course, in Computational Grid Systems is scheduling. However, scheduling jobs on the Grid is different from that for other systems for one or more of the following reasons: (1) The resources in the Grid are in different autonomous domains, while in other systems you find that resources within a single administrative domain (2). The resources are invariant while in the Grid the computing power available varies over the time. (3) The scheduler in the Grid does not have a single system image as the case for other parallel and distribute systems. The broad definition of the Grid [15] divulges the difficulty of scheduling on Grid if all Grid System characteristics and settings are to be considered; Grid defined in [15] as *"A type of parallel and distributed system that enables the sharing, selection, and aggregation of geographically distributed autonomous and heterogeneous resources dynamically at runtime depending on their availability, capability, performance, cost, and users' quality-of-service requirements".* Thus, in literature you find numerous scheduling algorithms and models for the Grid. Those algorithms and models can be categorized in different classes based on which of the Grid computing characteristics are considered. The subsets of the scheduling algorithms in Grids can be classified under one of the following categories:

- Local vs. Global
- Static vs. Dynamic
- Optimal vs. Suboptimal
- Approximate vs. Heuristic
- Distributed vs. Centralized
- Cooperative vs Independent

Obviously from the Grid definition and structure, the Grid scheduling cannot be local it has to be Global. Static scheduling assumes that all the resources available and the tasks to be scheduled are all known at the time we run the scheduling algorithm or model. An example of such algorithms found in [16-20]. Static algorithm is not fault-tolerant. That is, if a machine is not available (malfunction or the link to that machine is broken), we do not have a mechanism do re-distribute the load. To alleviate such problem, some algorithms consider rescheduling mechanisms [21]. The dynamic scheduling handles the situations where the available resources, tasks and the size of the tasks are not available at the time of scheduling. We refer to such scheduling as online scheduling. A good example of such algorithms found in [22]. In case of static mode, one can do an optimal assignment.

Otherwise, we should look for suboptimal solutions. Suboptimal solution can be either approximate or heuristic. Example of Heuristic algorithms can be found in [23]. Centralized schedulers suffer from single point failure problem and low performance under heavy load. Consequently, it is not scalable. If a distributed scheduling mechanism is adopted, the second important question is whether the schedulers reside in different nodes are working together or independently. An example of cooperative algorithm in Grid found in [24].

Another issue to consider in scheduling models is the application nature and objective. The subtasks of some applications are independent and others have some precedence relation. The later requires synchronization of those subtasks in different distributed processors in a Grid [25].

Since scheduling affects drastically the performance of Grids, one can find in the literature many articles discuss scheduling in Grid computing Systems. For example, The Resource CoAllocation scheduling algorithm [27] minimizes the execution time of a task; however, it suffers from high communication overhead. Optimal Resource Constraint (ORC) algorithm [28] and Job schedule Model based on the Grid [29] are also both suffer from high communication delay, though the first algorithm simplifies the allocation process and the later maximizes CPUs utilizations and the throughput. Some algorithms are fault-tolerant [30]; however, jobs have high waiting time. In the most of those algorithms, it is very hard to find optimal criteria. In [25], the computing power consumed by a schedule is set as a criterion of the schedule, and accordingly a performance limit is derived. The algorithm is not realistic since it deals only with coarse-grained applications, which implies that the communication delay is negligible.

Obtaining a closed form solution for the general problem where *n* subtasks be assigned to *N* geography distributed heterogeneous processors in a Grid such that optimum finish time is achieved talking into considering all dependencies among subtasks and the nature of the Grid is a very complex problem[26].

In this proposal we consider applications of divisible nature. A "divisible" load or Job (we will use Load and Job interchangeable) is a kind of load that is perfectly divisible by any number of available parallel machines (processors). Each fraction of the load can be communicated independently to a Grid (computing parallel machine) and executed independently as well. At the end, the results agglomerated at the originating node. The Divisible Load Theory (DLT) is a well established theory which seeks the optimal finish time of the load by scheduling a fraction of the load to each computing device (processors) in the heterogeneous parallel systems (here is the Grid) such that all subtasks on all the processors stop at the same time. DLT has emerged as a powerful tool to schedule divisible data-intensive applications on all sorts of distributed systems including Grids [31-35].

Recently, they were several attempts to exploit the DLT to model scheduling of arbitrarily divisible load on the Grid. Most of the attempts do not consider the communication time [12-14]. In [8] communication time is well thought-out but not in dividing the load so the transfer input time of the load was not part of the model. The communication and computation time are considered at the same time in [36]. However, the paper does not provide a closed form solution for the minimum finish time. The proposed work is aiming at alleviating the shortcoming of the previous work. So our objective is to come up with closed form solution for the minimum finish time of executing an arbitrarily divisible application on the Grid taking into considerations the communication time as well as the computation time simultaneously.

## Proposed work and Problem Definition

In this proposal we will attempt to model the Grid scheduling problem based on Divisible Load Theory taking into consideration the communication time as well as the computation time. We also consider that the load is available in multiple sources and will be executed in multiple heterogeneous worker nodes in the Grid. The load can be divided into **N** fractions where **N** is the number of heterogeneous worker nodes or simply nodes in the Grid. The analytical model should calculate the optimal fraction of the load that has to be assigned to each Node in the Grid such that the optimality criterion (Objective Function) is achieved. Our solution is based on "optimality Principle"[26]. To be more realistic to existing Grid Systems, we assume that the load is available in different sources and can be communicated to any of the **N** available Nodes. In calculating the optimal fraction of the load to be assigned to each Node (Sink) in the Grid, the heterogeneous speed of the links and the available computing power of the Nodes in Grid are modeled.

At this stage we consider the speed of the links and the speed of the nodes involved in computation of a given job is invariant during the execution period of the job. We believe this is not a realistic assumption though it is acceptable. In the future we plan to extend this work by releasing this constraint.

## Methods: The research approach

Since the objective is to come up with a scheduling analytical model and develop a closed form solution that shows the effect of the different parameters of the Grid system on the scheduling performance and on the system performance as well, I will use only the mathematical tools and analysis developed for DTL theory. May be a simulation is needed at later stages to confirm the validity of the analytical results.